\documentclass[preprint2]{aastex}

\shorttitle{Analysis of the motion of an extra-solar planet in a binary system}
\shortauthors{Plávalová et al.}
\usepackage{epsfig,graphicx,natbib,url,twoopt}

\usepackage{amssymb}	%pre špeciálne matematické symboly
\usepackage{amsmath}	%pre matematické prostredie rovníc
\usepackage{epsfig,graphicx,natbib,url,twoopt}
\usepackage[varg]{txfonts}
\usepackage[breaklinks]{hyperref}  
\begin{document}

\title{Analysis of the motion of an extra-solar planet
\\ in a binary system}

\author{Eva Pl\' avalov\' a}
\affil{Astronomical Institute, Slovak Academy of Science, Bratislava, Slovak Republic}
\email{plavala@slovanet.sk}

\and

\author{Nina A. Solovaya}
\affil{Sternberg State Astronomical Institute, Lomonosov Moscow State University, Moscow, Russia}
\email{solov@sai.msu.ru}

%%%%%%%%%%%%%%%%%%%%%%%%%%%%%%%%%%%%%%%%%%%%%%%%%%%%%%%%%%%%%%%%%%
\begin{abstract}
More than 10\% of extra-solar planets (EPs) orbit in a binary or multiple stellar system. We investigated the motion of planets revolving in binary systems in the frame of the particular case of the three body problem. We carried out an analysis of the motion an EP revolving in a binary system by following conditions; a) a planet in a binary system revolves around one of the components (parent star), b) the distance between the star`s components is greater than between the parent star and the orbiting planet (ratio of the semi-major axes is a small parameter), c) the mass of the planet is smaller than the mass of the stars, but is not negligible. The Hamiltonian of the system without short periodic terms was used. We expanded the Hamiltonian in terms of Legendre polynomial and truncated after the second order term depending on only one angular variable. In this case the solution of this system was obtained and the qualitative analysis of motion was produced. We have applied this theory to real EPs and compared to the numerical integration. Analyses of the possible regions of motion are presented. It is shown that the case of the stable and unstable motion of the EPs are possible. We applied our calculations to two binary systems hosting an EP and calculated the possible values for their unknown orbital elements.
\end{abstract}
%%%%%%%%%%%%%%%%%%%%%%%%%%%%%%%%%%%%%%%%%%%%%%%%%%%%%%%%%%%%%%%%%%
%%%%%%%%%%%%%%%%%%%%%%%%%%%%%%%%%%%%%%%%%%%%%%%%%%%%%%%%%%%%%%%%%%
\section{Introduction}
More than half of all main sequence stars reside in binary or multiple systems \citep{1991A&A...248..485D}. At least 10\% of the discovered extra-solar planets (EPs) have been observed to be orbiting binary or multiple stellar systems \citep{2011A&A...532A..79S}.  Orbits in multiple stellar systems can have a miscellaneous, in many cases unbelievable and improbable architecture. 

We targeted the binary stellar systems which are hosting EPs. For these systems \cite{1986A&A...167..379D} defined three different possibilities for stable planetary orbit: S-orbit (S-type) -- the Satellite-type orbit, where an EP orbits one of the stars; P-orbit (P-type) -- the Planet-type orbit, where an EP surrounds both stars; L-orbit (L-type) -- the Libration type orbit, where an EP is librating around one of the triangular Lagrangian points. We have focused on an S-type orbit in this paper. The EP's motion in such system was considered in the frame of the general three body problem.
%%%%%%%%%%%%%%%%%%%%%%%%%%%%%%%%%%%%%%%%%%%%%%%%%%%%%%%%%%%%%%%%%%
\section{Setting up the problem}
The motion of an EP is investigated in the frame of the general three-body problem, where the planet in the binary system revolves around one of the components (a parent star) and the mass of the planet is much smaller than the mass of the stars, but is not negligible. The distance between the binary components is greater than between the parent star and the orbiting planet (the ratio the semi-major axes is a small parameter). The motion is considered in the Jacobian coordinate system and the invariable plane is taken as the reference plane. We used the Delaunay canonical elements  $L_i$, $G_i$, $H_i$, $l_i$, $g_i$, and $h_i$ ($i$=1 for the planet's orbit, $i$=2 for the distant star's orbit). They can be expressed through the Keplerian elements as 
\begin{alignat}{3}
\label{delanay}
L_i&=\beta_i \sqrt{a_i}\,,  \quad 
G_i&=L_i  \sqrt{1-e_i^{2} }\,, \quad \; 
H_i&=G_{i}\,\cos{I_i}\,,
\nonumber\\
l_i&=M_i\,, \quad \quad \;
g_i&=\omega_i \,, \quad \qquad \qquad
h_i&=\Omega_i \,.
\nonumber\\
\end{alignat}
\\[-0.70cm] 
Where
\begin{align}
\label{beta}
   	\beta_1&=k\, \frac{m_0\,m_1}{ \sqrt{m_0+m_1}}=k{\mu_1}\,, 
\nonumber\\   	 
 	\beta_2&=k\, \frac{(m_0+m_1)\,m_2}{\sqrt{m_0+m_1+m_2}}=k{\mu_2}\,.   
\end{align}  
In the previous expression the notation has the usual meaning: $m_0$, $m_2$ –-- the masses of the stars, $m_1$ –-- the mass of the planet, $k$ -- the Gaussian constant, $a_i$ -- the semi-major axis, $e_i$ -- the eccentricity, $M_i$ -- the mean anomaly, $I_i$, $\omega_i$, $\Omega_i$ -- are the angular variables to the observation's plane,  and $g_i$ -- the argument  of the pericenter in the  invariable plane (this plane is perpendicular to the angular momentum of the system). The eccentricities of the star's and planet's orbits can have any value from $0<e_i<1$. 

In the general case the motion is defined by the masses of components and by the six initial values of the Keplerian elements of the planet and the distant star. The solution of a task using the Hamiltonian without short-periodic terms was obtained in the hyper elliptic integrals by the Hamilton-Jacobi method \citep{1988ASSL..140..243O}. The short-periodic terms are small and have no significant influence on the dynamic evolution of the system, the values of which are less than $\pm 10^{-3}$ which is less than the precision capabilities of the observations. The secular and essential long periodic terms are included into the Hamiltonian. Hamiltonian expanded in the terms of the Legendre polynomials and truncated after the second order terms carries the following form:
\begin{align}
\label{Hamiltonian}   
F&=\frac{\gamma_1}{2L_1^2}+\frac{\gamma_2}{2L_2^2}-
\nonumber\\
&-\frac{1}{16}\,\gamma_3\,\frac{L_1^4}{L_2^3 G_2^3} \left [\left(1-3q^2 \right)\,\left(5-3\eta^2 \right) - \right. 
\nonumber\\
&\left.-15\,\left(1-q^2\right)\,\left(1-\eta^2\right)\,cos{\left(2g_1\right)}\right]\,,
\end{align} 
where the coefficients $\gamma_1$,$\gamma_2$, and $\gamma_3$ depend on mass as follows:
\begin{eqnarray}
\label{gamma}
\gamma_1=\frac{\beta_1^4}{\mu_1}\,,\quad 
\gamma_2=\frac{\beta_2^4}{\mu_2}\,, \quad 
\gamma_3=k^2\,\mu_1\,\mu_2\,\frac{\beta_2^6}{\beta_1^4}\,,  
\end{eqnarray} 
the parameter
\begin{eqnarray} 
\label{eta}
\eta=\sqrt{1-e_1^2} \, ,  
\end{eqnarray}
and the cosine of the angle between the plane of the planet's orbit and the plane of the distant star's orbit
\begin{eqnarray} 
\label{cos}
q=\frac{c^2-G_1^2-G_2^2}{2\,G_1\, G_2}\, .  
\end{eqnarray}
Where $c$ is the constant of the angular momentum of the system, and
\begin{equation}
c=\sqrt{C_x^2+C_y^2+C_z^2} \, ,
\end{equation}
where
\begin{align}
\label{definiciaLapplace}
C_x&=\sum_{i=1}^2 \beta_i \sqrt{{a_i}\left( 1-e_i^2\right) }\sin I_i  \sin {\Omega_i} \, ,
\nonumber\\
C_y&=\sum_{i=1}^2 \beta_i \sqrt{{a_i}\left( 1-e_i^2\right) }\sin I_i  \cos {\Omega_i} \, ,
\nonumber\\
C_z&=\sum_{i=1}^2 \beta_i \sqrt{{a_i}\left( 1-e_i^2\right) }\cos I_i  \, . 
\end{align}

The Hamilton's equation truncated after the second order terms depends on the one angular variable $g_1$, only. It is necessary to integrate two equations from the canonical system of the differential equations.
\begin{equation}
\label{canonical}
\frac{dG_1}{dt}=\frac{\partial F}{\partial g_1} \, ,
\qquad   \qquad
\frac{dg_1}{dt}=-\frac{\partial F}{\partial G_1} \, .
\end{equation}
Then the other equations will be as follows:
\begin{alignat}{3}
\label{equationmotion}
\frac{dL_i}{dt}&=0 \, , \qquad \qquad \quad
\frac{dl_i}{dt}&=-\frac{\partial F}{\partial L_i} \, ,
\nonumber\\
\frac{dG_2}{dt}&=0 \, , \qquad \qquad \quad \frac{dg_2}{dt}&=-\frac{\partial F}{\partial G_2} \, , 
\nonumber\\
\frac{dc}{dt}&=0 \, , \qquad \qquad \quad
\frac{dh}{dt}&=-\frac{\partial F}{\partial c} \,,
\end{alignat}
where $ h=h_1$, and $h_2=h+180^\circ$. 
If system (\ref{canonical}) is integrated, the angular variables $l_i$, $g_2$, and $h$ will be defined from (\ref{equationmotion}) in quadratures. For the solution of system (\ref{canonical}), we used the Hamilton-Jacobi method. According to this method, we must find the complete integral
\begin{equation}
\label{W}
W=W\left( l_1, l_2, g_1, g_2, h, A_1, A_2, A_3, A_4, A_5\right) 
\end{equation}
of the equation in the partial derivations
\begin{equation}
\label{dW/dt}
\frac{\partial W}{\partial t}+\Phi\left( g_1,\frac{\partial W}{\partial l_1}, \frac{\partial W}{\partial l_2}, \frac{\partial W}{\partial g_1}, \frac{\partial W}{\partial g_2}, \frac{\partial W}{\partial h}\right) =0 \, ,
\end{equation}
where $\Phi=-F$ and $A_1$, $A_2$, $A_3$, $A_4$, $A_5$ are the arbitrary constants. The general solution of the system is presented as follows:
\begin{alignat}{2}
\label{generalsolution}
L_i&=\frac{\partial W}{\partial l_i} \, , \qquad \qquad \quad B_i&=\frac{\partial W}{\partial A_i} \, ,
\nonumber\\
G_i&=\frac{\partial W}{\partial g_i} \, , \qquad \qquad \quad c&=\frac{\partial W}{\partial h} \, .
\end{alignat}
So, as the Hamiltonian does not depend on $l_i$, $g_2$, and $h$, the integral (\ref{W}) is as follows:
\begin{equation}
\label{WW}
W=\varepsilon\left( t-t_0\right) +A_1l_1+A_2l_2+A_4g_2+A_5h+W_1(g_1) \, ,
\end{equation}
where $\varepsilon$ is a new constant and $W_1(g_1)$ is a new function which we must define. When we substituted the function $W_1(g_1)$, defined by (\ref{WW}) to (\ref{dW/dt}), we found the equation  to satisfy the function $W_1$:
\begin{equation}
\label{ee}
\varepsilon+\Phi\left( g_1, A_1, A_2, W'_1(g_1), A_4,A_5\right) =0 \, ,
\end{equation}
where $W'(g_1)$ is the derivation. Using equation (\ref{Hamiltonian}), we rewro\-te previous equation to
\begin{eqnarray}
\label{varepsilon}
\varepsilon = \frac{\gamma_1}{2A_1^2}+\frac{\gamma_2}{2A_2^2}-\frac{1}{16}\gamma_3 \frac{A_1^4}{A_2^3 A_4^3}A_3 \, ,
\end{eqnarray}
where the constant $A_3$ has the following form:
\begin{align}
\label{A3}
A_3&=\left[ 1-\frac{3}{4}\frac{\left( A_5^2-A_4^2-W'{_1^2}\right) ^2 }{A_4^2 W'{_1^2}} \right]\left( 5-3\frac{W'{_1^2}}{A_1^2}\right)- 
\nonumber\\
&-15\left[ 1-\frac{1}{4} \frac{\left( A_5^2-A_4^2-W'{_1^2}\right) ^2 }{A_4^2 W'{_1^2}} \right]\times
\nonumber\\
&\times\left( 1- \frac{W'{_1^2}}{A_1^2} \right)\cos \left( 2g_1\right) \, .
\end{align}
The equation (\ref{A3}) is an ordinary differential equation of the first order. If the solution is found then the general solution of the canonical system is:
\newpage
\par\noindent
\begin{alignat}{2}
\label{solution}
L_1&=A_1 \, , \qquad \quad  B_1&=\frac{\partial \varepsilon}{\partial A_1}\left( t-t_0\right) +l_1+\frac{\partial W_1}{\partial A_1} \, ,
\nonumber\\
L_2&=A_2 \, , \qquad \quad B_2&=\frac{\partial \varepsilon}{\partial A_2}\left( t-t_0\right) +l_2+\frac{\partial W_1}{\partial A_2} \, ,
\nonumber\\
G_1&=\frac{\partial W_1}{\partial g_1} \, ,  \qquad  B_3&=\frac{\partial \varepsilon}{\partial A_3}\left( t-t_0\right) +\frac{\partial W_1}{\partial A_3}\, , \qquad 
\nonumber\\
G_2&=A_4 \, , \qquad \quad B_4&=\frac{\partial \varepsilon}{\partial A_4}\left( t-t_0\right) +g_2+\frac{\partial W_1}{\partial A_4} \, ,
\nonumber\\
c&=A_5 \, , \qquad \quad B_5&=\frac{\partial \varepsilon}{\partial A_5}\left( t-t_0\right) +h+\frac{\partial W_1}{\partial A_5} \, .
\end{alignat}
In the third line of the system (\ref{solution}), one can see the dependence between $\xi$ and time $t$ where,
\begin{equation}
\label{xi}
\xi=\frac{W_1'^2}{A_1^2}=\frac{G_1^2}{A_1^2} \, .
\end{equation}
After the differentiation and the algebraic operations, we obtained the following equation  connecting $\xi$ and $t$: 
\begin{equation}
\label{orlov}
\frac{1}{12} \overline{G}_2^2 \int_{\xi_1}^{\xi} \frac{{\rm d}\xi}{\sqrt{\Delta}} = \frac{B_3}{A_1}+\frac{1}{16} \frac{\gamma m^{\prime\prime^2}}{\left( 1-e_2^2 \right)^{\frac{3}{2}}} n_1 \left( t - t_0 \right).
\end{equation}
Where \mbox{$m^{\prime \prime}= {n_2}/{n_1} $} and $n_1$, $ n_2$ -- the mean motions of the planet and the distant star, $\displaystyle{\gamma={m_2}/({m_0+m_1+m_2})}$, while $A_1$ and $B_3$ are the constants of integration.

In this approximation, we obtained an exact solution. We used this solution as an intermediary orbit for the planet's motion in which the second order  perturbations are included. The expressions for $a_1$ and $a_2$ do not contain secular terms and hence are restricted in their time evolution. 

The change of the eccentricity of a planet is described by the expression $e_1=\sqrt{1-\xi}$. In the case when the eccentricity of a planet's orbit can change to almost as much as 1, the planet's pericenter is near to or beyond the Roche limit, and large perturbations and tidal forces lead to the destruction of the planet. Such a system will be dynamically unstable.

On the left side of the equation (\ref{orlov}) under the integral in the denominator is the square root from the polynomial of the fifth order. This polynomial can be presented as the product of the  two polynomials –  the second and the third  orders,  $\Delta=f_2(\xi)f_3(\xi)$, where $\xi=1-e_1^2$, and $e_1$ is the eccentricity of the planet's orbit. The determination of the regions of the motion are possible when the roots of the equations $f_2(\xi)=0$, $f_3(\xi)=0$ are found and the signs of the functions defined. 
%%%%%%%%%%%%%%%%%%%%%%%%%%%%%%%%%%%%%%%%%%%%%%%%%%%%%%%%%%%%%%%%%%
\section{Equations of the second  and third order}
We investigated the roots of the equations \mbox{$f_2(\xi)=0$} and \mbox{$f_3(\xi)=0$} with following forms:
\begin{align}
\label{f2}
	f_2(\xi)&=\xi^2-2 \left(\overline{c}^2+3\overline{G}_2^2  \right)\xi+\left(\overline{c}^2-\overline{G}_2^2\right)^2+
\nonumber\\	
	&+\frac{2}{3}\left(10+A_3\right)\overline{G}_2^2
\nonumber\\	
\end{align}
and
\begin{align} 
\label{f3}
	f_3 (\xi)&=\xi^3-\left(2\overline{c}^2+\overline{G}_2^2+\frac{5}{4}\right)\xi^2+
\nonumber\\	
	&+\left[ \frac{5}{2}\left(\overline{c}^2+\overline{G}_2^2\right)+ \left(\overline{c}^2-\overline{G}_2^2\right)^2- \right. 
\nonumber\\
&\left.-\frac{1}{6}\,\overline{G}_2^2\left(10+A_3\right) \right] \xi-\frac{5}{4} \left(\overline{c}^2-\overline{G}_2^2\right)^2\,.	
\nonumber\\	
\end{align}
Where
\begin{eqnarray}
\overline{c}=\frac{c}{L_1}\,, \qquad \qquad \overline{G}_2>1\,,
\nonumber\\
 \overline{G}_2=\frac{G_2}{L_1}=\frac{\beta_2}{\beta_1}\sqrt{\frac{a_2\left( 1-e_2^2\right)}{a_1}}   \,, 
\end{eqnarray}
and
\begin{eqnarray}
\label{a3}
	A_3=2-6\eta_0^2 q_0^2-6\left(1-\eta_0^2\right)\times
	\nonumber\\
	\times\left[2-5\left(1-q_0^2\right)sin^2 {g_{10}}\right]\,.  
\end{eqnarray}
We note that 
\begin{eqnarray}
	\eta_0=\sqrt{1-e_{10}^2}\,,
\end{eqnarray}
where, $e_{10}$ is the initial value of the eccentricity of the planet, $q_0$ is the initial value of the cosine of the mutual inclination between the orbits of the planet and the distant star and $g_{10}$ is the initial value of the argument of the pericenter of the planet's orbit in the invariable plane.

So the orbit is assumed to be elliptic, $0 <e_1 <1$, then  $0 < \xi <1$. Value of $\xi=1$ corresponds to a circular motion. Such a case was not considered in this paper.

For the determination of  the boundaries of the regions of the possible motion  it is necessary to find the roots of the equations (\ref{f2}) and (\ref{f3}) and to define the signs of the function in the interval between the roots.
%%%%%%%%%%%%%%%%%%%%%%%%%%%%%%%%%%%%%%%
\subsection{The equation of the second order}
We rewrote equation of the second order (\ref{f2}) in the following form:
\begin{align}
\label{f2rewr}
	f_2(\xi)&=
\xi^2-2 \left(\overline{c}^2+3\overline{G}_2^2  \right)\xi+
\nonumber\\
&+\left[-1+2 \left(\overline{c}^2+3\overline{G}_2^2  \right)+\frac{2}{3}\overline{G}_2^2\overline{h}\right]\, ,
\end{align}
where $\overline{h}$ is difference between values $A_3$ and $A_{3crit}$
\begin{eqnarray}
\label{h}
 \overline{h}=A_3-A_{3crit} \, .
\end{eqnarray}
Then
\begin{align}
2\overline{G}_2^2\overline{h}&=3(1-\eta^2)
\left[1-8\overline{G}_2^2-4\overline{G}_2 q\eta-\eta^2+ \right.
\nonumber \\
	&\left. +20\overline{G}_2^2\left(1-q^2\right)sin^2g_1\right]\,.
\end{align}
We named the meaning of the constant $A_3$ as $A_{3crit}$ for which the square equation has the root equal to one. We found $A_{3crit}$ when  $f_2(\xi)=0$ and $\xi=1$.
We note the roots of the equation (\ref{f2rewr}) as $\epsilon_1$ and $\epsilon_2$, and $\epsilon_1<\epsilon_2$. The coefficient by $\xi$ is always less than zero. This equation has no negative roots and the free term is also greater then 0. The roots are
\begin{align}
\label{epsilon12}
\epsilon_{1,2}&=\left(\overline{c}^2+3\overline{G}_2^2\right)\mp
\nonumber\\
&\pm\sqrt{\left(\overline{c}^2+3\overline{G}_2^2+1\right)^2-\frac{2}{3}\overline{G}_2^2\overline{h}} \,,
\end{align}
and $\epsilon_2>>1$ always. Derivation $(d\epsilon_1)/(d\overline{h})$ is always positive. So  $\epsilon_1$ is a growing function of $\overline{h}$. When  $\overline{h}=0$ then $\epsilon_1=1$, we conclude for $\overline{h}<0$ then $\epsilon_1<1$ , and  for $\overline{h}>0$ then $\epsilon_1>1$. For the definition of the branch of the parabola we found the extreme of the function in the point which is valid $\overline{\xi}=\overline{c}^2+3\overline{G}_2^2$. The value $f_2(\overline\xi)<0$ and the branches of this parabola are directed up.
%%%%%%%%%%%%%%%%%%%%%%%%%%%%%%%%%%%%%%%
\subsection{Equation of the third order}
%_______________________________________________
\begin{figure}
   \centering 
     \includegraphics[width=5.5cm]{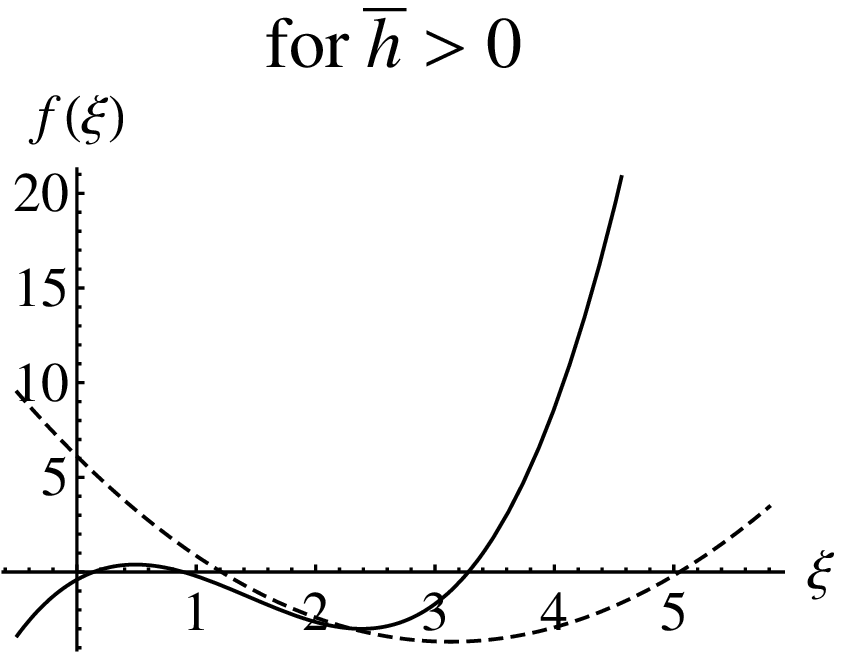}
     \caption{Behaviour of the functions $f_2(\xi)$ (dashed line) and $f_3(\xi)$ (solid line) for $\overline{h}>0$.}
\label{graph_hplus}
 \end{figure}
\begin{figure}
   \centering 
   \includegraphics[width=5.5cm]{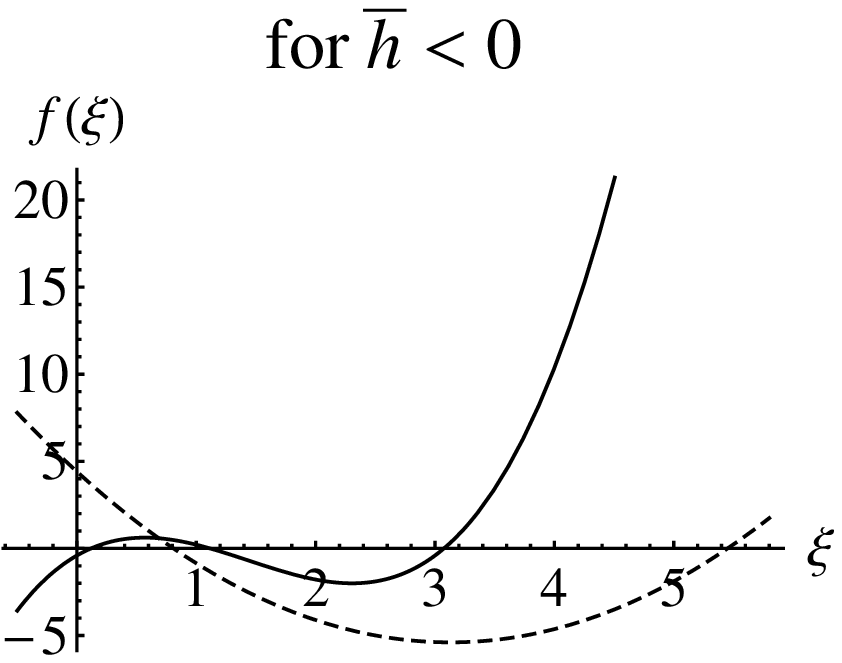}
         \caption{Behaviour of the functions $f_2(\xi)$ (dashed line) and $f_3(\xi)$ (solid line) for $\overline{h}<0$.}
\label{graph_hminus}
\end{figure}

%________________________________________________
We rewrote equation (\ref{f3}) using $\overline{h}$ in the form:
\begin{align}
\label{f3prepis}
f_3 (\xi)&=\xi^3-\frac{1}{4}\left(5+8\overline{c}^2+4\overline{G}_2^2\right)\xi^2+
\nonumber\\
&+\left[\frac{1}{4}+2\overline{c}^2+\overline{G}_2^2+\frac{5}{4}\left(\overline{c}^2-\overline{G}_2^2\right)^2- \right.
\nonumber \\ 
	&\left. -\frac{1}{6}\overline{G}_2^2\overline{h}\right]\xi-\frac{5}{4}\left(\overline{c}^2-\overline{G}_2^2\right)^2 \,.
\end{align}
Then multiplied this equation by 4 and identified in its left part the term $\xi f_2(\xi)$. It is possible to rewrite equation (\ref{f3prepis}) to take the following form:
\begin{align}
\label{f3root}
f_3(\xi)&=5\left(1-\xi\right)\left[\left(\overline{c}+\overline{G}_2\right)^2-\xi \right]
\left[\xi-\left(\overline{c}-\overline{G}_2\right)^2 \right] -
\nonumber\\
 &-\xi f_2(\xi)\,.
\end{align}
For the determination of the qualitative characteristics of the motion, it is necessary to define how the roots of the equation of the third order $f_3(\xi)=0$ are located in relation to the points of the axes:
\begin{eqnarray}
\xi=0, \qquad \xi=\epsilon_1, \qquad \xi=1, \qquad  \xi=\epsilon_2.
\end{eqnarray}
\begin{enumerate}
\item
If $\overline{h}<0$ then we have:
\begin{enumerate}
\item
$\xi=0$,  \, \qquad $f_3(0)<0$.
\item
 $\xi=\epsilon_1$, \qquad $f_3(\epsilon_1)>0$.
\item
 $\xi=1$, \, \qquad $f_3(1)>0$.
\item
 $\xi=\epsilon_2$, \qquad $f_3(\epsilon_2)>0$.
\end{enumerate}
From Sturm theorem \citep[e.g.][]{dorrie1965100} it follows that, between 0 and $\epsilon_1$ lay at least one root of the third order equation.  
\item
If $\overline{h}>0$ then we have:
\begin{enumerate}
\item
$\xi=0$ , \,  \qquad $f_3(0)\leqslant 0$.
\item
$\xi=1$ ,  \, \qquad $f_3(1)<0$.
\item
$\xi=\epsilon_1$ , \qquad $f_3(\epsilon_1)<0$.
\item
$\xi=\epsilon_2$ , \qquad $f_3(\epsilon_2)>0$.
\end{enumerate}
\end{enumerate}
The behaviours of the roots of the second and third order equations depend on the sign of $\overline{h}$ and valid:
\begin{enumerate}
\item
for $\overline{h}<0$ is valid $0\leqslant \epsilon_3 < \epsilon_1 < 1 < \epsilon_4 < \epsilon_5 < \epsilon_2$.
\item
for $\overline{h}>0$ is valid $0\leqslant \epsilon_3 < \epsilon_4 < 1< \epsilon_1 < \epsilon_5 < \epsilon_2$.
\end{enumerate}
We have two roots which are less than 1, therefore we identified the roots in ascending order as $\xi_1,  \xi_2, \xi_3, \xi_4, \xi_5$. This means that $e_1$ can change in limits from $e_{1min}=1-\xi_2$ to $e_{1max}=1-\xi_1$.
We showed the behaviour of these functions for the hypothetical value $\overline{c}$, $\overline{G}_2$, and $\overline{h}$ on the Figure \ref{graph_hplus} for $\overline{h}<0$ and on Figure \ref{graph_hminus} for $\overline{h}>0$.
%_______________________________________________
%_________________________________________________
%%%%%%%%%%%%%%%%%%%%%%%%%%%%%%%%%%%%%%%%%%%%%%%%%%%%%%%%%%%%%
\section {The investigation of the value $\mathbf{\overline{h}}$}
Consider the value $\overline{h}$, which has the following form:
\begin{align}
\label{hh}
\overline{h}&=\frac{3(1-\eta^2)}{2\overline{G}_2^2}
\left[1-8\overline{G}_2^2-4\overline{G}_2 q\eta-\eta^2+ \right.
\nonumber \\
	&\left. +20\overline{G}_2^2\left(1-q^2\right)sin^2g_1\right]\,.
\end{align}
When the value $g_1=0$ then value $\overline{h}$ reaches minimum
\begin{eqnarray}
\overline{h}_{min}=\frac{3(1-\eta^2)}{2\overline{G}_2^2}
\left[1-8\overline{G}_2^2-4\overline{G}_2 q\eta-\eta^2\right]\,.
\end{eqnarray}
The value of the $\overline{h}_{min}$ is always negative. In this case we have the two roots valued less than 1, the one root of the equation of the third order and one root of the square equation. 
If $g_1=\pi/2$ then $\overline{h}$ has a maximum value
\begin{align}
\overline{h}_{max}&=\frac{3(1-\eta^2)}{2\overline{G}_2^2}
\left[1-\eta^2-4\overline{G}_2 q\eta+ \right.
\nonumber \\
	&\left.+12\overline{G}_2^2-20\overline{G}_2^2 q^2\right].
\end{align}
The value of $\overline{h}_{max}$ can either be negative or positive.\\
For the case $g_1=\pi/2$ we rewrote equations (\ref{f2}) and (\ref{f3}) in the following form:
\begin{align}
\label{f2prepis1}
f_2(\xi)&=\left( \xi-\eta_0^2\right)\left( \xi-\eta_0^2-4\overline{G}_2\eta_0q_0+8\overline{G}_2^2\right)+ 
\nonumber\\
&+4\overline{G}_2^2\left( 1-5q_0^2\right)\left( 1-\eta_0^2\right)+16\overline{G}_2^2 
\end{align}
and
\begin{align}
\label{f3prepis1}
f_3(\xi)&=\left( \xi-\eta_0^2\right) \left[ \left( \xi-\frac{5}{4}\right) \left( \xi-\eta_0^2\right)-
\right.
\nonumber \\ 
&\left. 
-3\overline{G}_2^2\xi+\overline{G}_2\eta_0q_0\left( 5-4\xi\right) +5\overline{G}_2^2q_0^2\right] \,.
\end{align}
When one of the roots of the equation (\ref{f3prepis1}) carries the value of $\xi=\eta_{0}^2$, the other roots have to satisfy the equation
\begin{align}
\label{xisarovnaetanula}
& \left({\xi-\frac{5}{4}}\right)\left( {\xi-\eta_{0}^2}\right)-
\nonumber\\
&-3\,\overline G_{2}^2\, \xi+\overline G_{2}\, \eta_{0}\, q_{0} \left(5-4\, \xi\right)+ 5\, \overline G_{2}^2\, q_{0}^2 =0 \, .
\end{align}
For $\overline h_{max}$  the initial value $\eta^2=\eta_{0}^2$ is one of the boundary limits for change in the value of $\xi$. We should establish that \mbox{$\xi=\eta_{0}^2$} is the least root and the second root has a value less than $1$. We therefore substitute the value $\xi=\eta_{0}^2$ in (\ref{xisarovnaetanula}). If the obtained expression is negative, then $\eta_{0}^2$ should become the second root in equation (\ref{f3prepis1}), if positive -- then  it should become the least root.
The left part of the equation (\ref{xisarovnaetanula}) for $\xi=\eta_{0}^2$ is
\begin{eqnarray}
\label{expression}
2\left[ 5q_0^2-3\eta_ 0^2+\frac{1}{\overline{G}_2}\eta_ 0 q_0\left( 5-4\eta_ 0 ^2\right) \right]  \, .
\end{eqnarray}
Equating this expression to zero and solving the obtained equation concerning $q_0$, we find the value of $q_0$ of which $\eta_0^2$ is the root of expression(\ref{expression}). So
\begin{eqnarray}
\label{q}
q=\frac{\eta_ 0\left[ 4\eta_ 0^2-5 \mp \sqrt{60\overline{G}_2^2+\left( 5-4\eta_ 0^2\right)^2}\right]} {10\overline{G}_2} \, .
\end{eqnarray}
We denote the meaning of roots (\ref{q}) as $q_{01}$ for the minus sign before the root term and the $q_{02}$ for plus sign. If the value of $q_{0}$ lays inside the region of $q_{01} < q_{0}  < q_{02}$, expression (\ref{expression}) is negative, and the value of the root lays between the other two roots. If either of the conditions $q_0 < q_{01}$ or $q_0 > q_{02}$ is valid, then $\xi=\eta_{0}^2$ is the least root of equation (\ref{xisarovnaetanula}). 

In this case $e_{1_{\rm max}}=\sqrt{1-\eta_0^2}$ and the maximum value of the eccentricity of the planet's orbit can not exceed the initial value of eccentricity. The orbit of the EP may be dynamical stable.

When the starting value of the cosine of the mutual inclination is  $q_0=-\eta_0/2\overline{G}_2$, then from Eg.(\ref{cos}) $\overline{c}^2-\overline{G}_2^2\approx 0$ and $e_1\rightarrow 1$.
%%%%%%%%%%%%%%%%%%%%%%%%%%%%%%%%%%%%%%%%%%%%%%%%%%%%%%%%%%%%%%%%%
\section{Application of the theory on real extra-solar planets}
The knowledge of the six pairs of Keplerian elements of the orbits of the system allow us to investigate the character of the evolution of the planet's orbit and the possible conditions of stability. They may be presented  by the orbital parameters, which we obtain from the analytical theory. They are -- the angle of the mutual inclination between orbits of the planet and star, the angular moment of the star and the maximum value of the eccentricity of the planet's orbit. The growth of the eccentricity of the orbit could lead to the destruction of the orbit in pericenter from tidal forces.

In the catalogue of EPs the data for the longitude of the ascending node and the value of inclination, is generally absent.  The existing observational techniques do not allow estimation of these two elements unambiguously.
 Our theory allows us to define a range of possible values for these unknown elements, by which the planet's eccentricity does not increase. In our application of the theory, we have selected two binary stellar systems, with a planet revolving around one of the components. The first is system 16 Cyg with an interesting planet 16 Cyg Bb, which has the argument of pericenter close to $90^{\circ}$. The second system is HD19994.
%%%%%%%%%%%%%%%%%%%%%%%%%%%%%%%%%%%%%%%%%%%%%%%%%%%%%%%%%%%
\subsection{16 Cyg}
%_________________________________________________
\begin{table*}[t]
\centering
\caption{Initial orbital elements of the system 16 Cyg.} 
\begin{tabular}{l c c } % 3 columns 
\hline 
  &star 16 Cyg A&planet 16 Cyg Bb\\
\hline \hline 
Mass [$M_{Sun}$]& 1.53 & \\
Mass [$M_{Jup}\times \sin I_1$]&&1.68$\pm$0.07 \\
Semi-major axis [AU]& 754.53 & 1.68$\pm$0.07\\
Eccentricity&0.863& 0.689$\pm$0.011 \\
Inclination&135.44$^\circ$ &\\
Ascending node&313.44$^\circ$&\\
Argument of pericenter&26.6$^\circ$&83.4$\pm$2.1$^\circ$\\
Period&13512.7 years&799.5 days\\
\hline 
\end{tabular}
\tablecomments{For the calculations we copied the orbital elements for 16 Cyg A used by \citet{1999PASP..111..321H} and for the planet we used elements from \citet{2011A&A...532A..79S}. We used the planet's semi-major axis value 1.6923 AU for the calculations. The mass of the parent star 16 Cyg B is $1.01\pm0.04 M_{Sun}$. } 
\label{table:1}
\end{table*}
%_________________________________________________
\begin{table*}
\centering
\caption{Our proposal for possible orbital elements for planet 16 Cyg Bb.}
\begin{tabular}{l c c} % 3 columns 
\hline 
  &Prograde orbit&Retrograde orbit\\
\hline \hline 
Mass [$M_{Jup}$]&2.38$\pm$0.04&2.38$\pm$0.04\\
Semi-major axis [AU]&1.693&1.693\\
Eccentricity&0.689$\pm$0.011&0.689$\pm$0.011 \\
Inclination&45$^\circ \pm$1$^\circ$&135$^\circ \pm$1$^\circ$ \\
Argument of pericenter&83.4$\pm$2.1$^\circ$&83.4$\pm$2.1$^\circ$\\
Period [day]&799.5&799.5\\
\hline 
\end{tabular}
\label{table:2} 
\end{table*}
%___________________________________________________
16 Cyg A (HD 186408) and 16 Cyg B (HD 186427) are both members of a well-known wide binary system with stars spectral types G1.5V and G3V. For the calculations, we used orbital elements used by \citet{1999PASP..111..321H}, which were published in the Sixth Catalogue of Orbits of Visual Binary Stars \citep{2001AAS...199.0609M}. The semi-major axis $a_2$ was calculated to be 754.53 AU using the values 21.41 pc \citep{1998A&A...336..942F} for the distance and $35.242 ^{\prime\prime}$ for the angular separation. We made a revision calculation for this semi-major axis using the third Keplerian law and the value of 13512.7 year for a period $P_2$ \citep{2001AAS...199.0609M}. We reached a value of $a_2 =774.06$ AU. There is an obvious difficulty in determining orbital parameters for periods greater than 10,000 years. Only a small fraction of the orbit has transpired since astrometric measurements were first obtained in the early 1800s. So, we used the value 754.53 AU, derived from the distance, for the following calculations because we considered the determination period less precise than the determination of the values for angular separation and distance. 

The secondary star in the 16 Cyg binary system, 16 Cyg B (parent star), with a mass of 1.01$\pm$0.04 $M_{Sun}$ \citep{1998A&A...336..942F}, is known to host a giant gas planet 16 Cyg Bb with taxonomy class 2J0.2W7 \citep{2012AsBio..12..361P}. The EP has a minimum mass $M_{Jup}\times \sin I_1=1.68\pm 0.07 M_{Jup}$ in an orbit with high eccentricity $e_1=0.689$ \citep{2007ApJ...654..625W}. The initial conditions for investigating this planet, we took from The Extrasolar Planets Encyclopaedia \citep{2011A&A...532A..79S}. The values of the inclination and ascending note for the planet are unknown. In Table \ref{table:1} are shown the orbital elements for the planet and distant star.

We made a revision calculation for the semi-major axis of the planet, using the third Keplerian law and values for the orbital period of the planet, which is 799.5 days. 
It is known the minimum mass of the planet $m_1= 1.68$ M$_{\rm Jup}$. We varied the elements $I_1$ from $0^\circ$ to $180^\circ$, $1^\circ$ at a time  and $\Omega_1$ from $0^\circ$ to $360^\circ$, also for each degree. We recognized the value of the planet's inclination $I_1$ for which the maximum value of the planet's eccentricity $e_1$ is close to the initial value 0.689. 

According to our calculation we received the value $1.693$ AU
 for the semi-major axis, $2.38 M_{Jup}$ for the mass, and $45^\circ$ or $135^\circ$ for the inclination of the planet. These values are presented in Table \ref{table:2} and were used for our next calculations.

The value of the ascending node of the planet's orbit $\Omega_1$ is unknown. But the orientation of the orbit is defined by the node, therefore the different ways of the dynamical evolution are possible. We varied the ascending node of the planet  from $0^\circ$ to $360^\circ$, by step $1^\circ$. We recognized the region of the values of $\Omega_1$ for which the planet's orbit would be stable or unstable.  
%%%%%%%%%%%%%%%%%%%%%%%%%%%%%%%%%
\begin{figure}[t]
   \centering 
     \includegraphics[width=7.5cm]{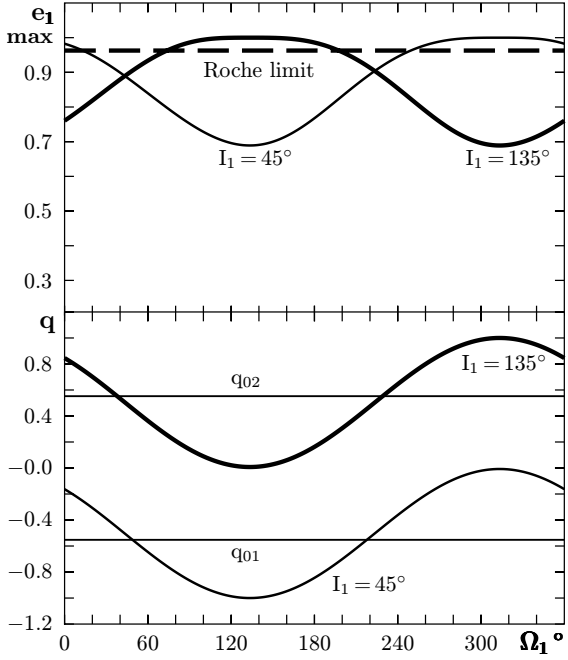}
     \caption{The planet 16 Cyg Bb. The evolution of the maximum value of the planet's eccentricity $e_1$ and the cosine of the mutual inclination between the planet's orbit and the distant star's orbit $q$ versus the ascending node of the planet $\Omega_1$ for the prograde, $I_1=45^\circ$, and retrograde, $I_1=135^\circ$,  planet orbits. The Roche limit  is plotted by a dashed line.}
\label{fig3}
 \end{figure}
 %%%%%%%%%%%%%%%%%%%%%%%%%%%%%%%%%

For the stability criteria, we used the Roche limit. This is the minimum distance which a planet can approach its parent star without being torn apart by tidal forces. For the calculation of the Roche limit $d_R$ in our paper, we used the equation published by \cite{1983ApJ...268..368E}.

\begin{equation}
d_R =\frac{0.49\, \mu^\frac{2}{3}}{0.6\, \mu^\frac{2}{3}+\ln {\left( 1 + \mu^\frac{1}{3} \right) }}\,,
\end{equation}
where
$\mu = m_1 / m_0$.

The results of our calculations are presented in Figure \ref{fig3}. and \ref{fig5}. If  the ascending node is $\Omega_1 \in \left( 249^\circ , 374^\circ \right) $ for prograde motion, $I_1=45^\circ$, and $\Omega_1 \in \left( 73^\circ , 198^\circ \right) $ for retrograde motion, $I_1=135^\circ$, the planet reaches the Roche limit in its pericentre ($d_R =0.063$ AU) with an eccentricity of $e_1 \ge 0.963$. In such a pericenter, large perturbations and tidal forces drastically effect the planet and lead to its destruction. 

For the values $I_1=45^{\circ}$  and $\Omega_1=135^{\circ}$ the maximum and minimum values of eccentricity $e_1$ are equal to the initial value $e_1 = 0.689$. In this case, $q_0 =-0.999 < q_{01}=-0.561$. 
With these elements, the planet's orbit would be stable. 

When the eccentricity of the planetary orbit grows close to the Roche limit, large perturbations in the planet's pericenter are affected. In such cases, measurements based on point mass bodies can be inaccurate, rather a dynamic theory for real dimensional bodies must be used.
%%%%%%%%%%%%%%%%%%%
\begin{figure}[t]
   \centering 
     \includegraphics[width=7.5cm]{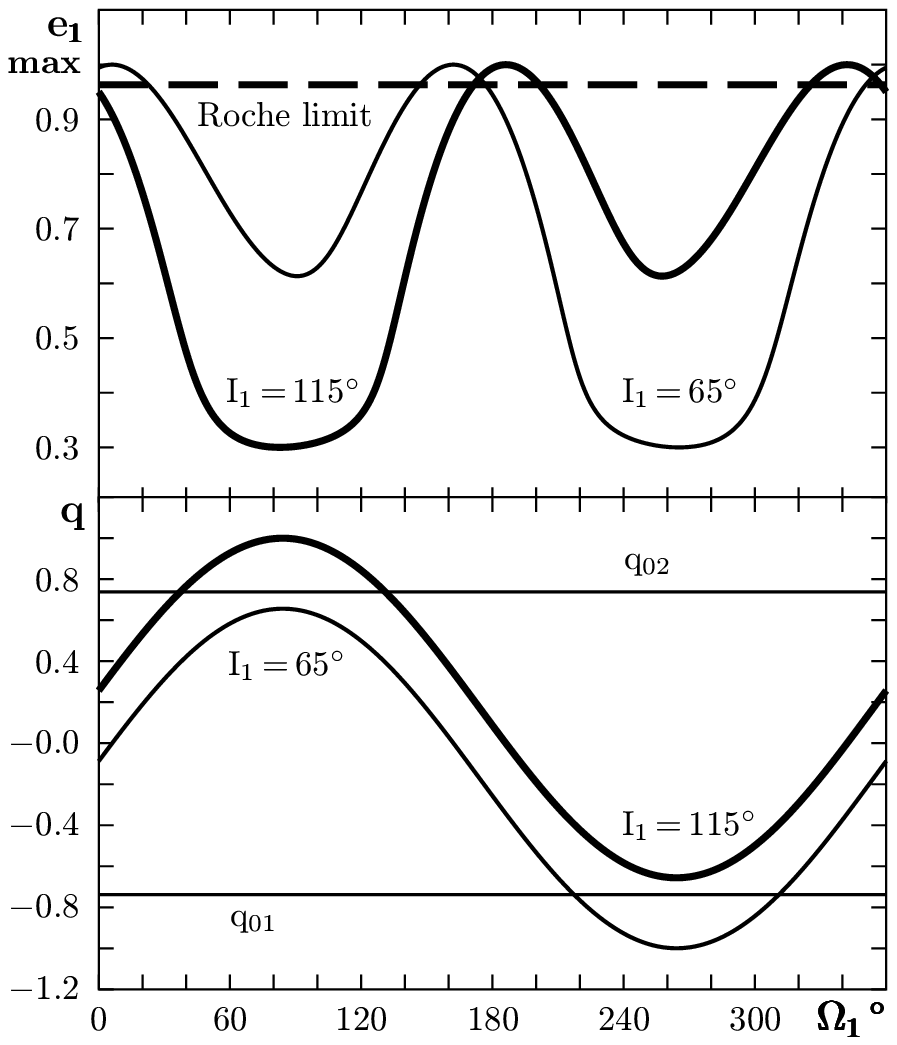}
     \caption{The planet HD19994 b. The evolution of the maximum value of the planet's eccentricity $e_1$ and the cosine of the mutual inclination between the planet's orbit and the distant star's orbit $q$ versus the ascending node of the planet $\Omega_1$ for the prograde, $I_1=65^\circ$, and retrograde, $I_1=115^\circ$,  planet orbits. The Roche limit  is plotted by a dashed line.}
\label{fig4}
 \end{figure}

%%%%%%%%%%%%%%%%%%%%%%%%%%%%%%%%%%%%%
\subsection{HD 19994}

\begin{table*}[t]
\centering 
\caption{Initial orbital elements of the system HD19994.} 
\begin{tabular}{l c c } % 3 columns 
\hline 
 &  star HD19994&planet HD19994 b\\
\hline \hline 
Mass [$M_{Sun}$]& 0.37& \\ 
Mass [$M_{Jup}\times \sin I_1$]&&1.68\\
Semi-major axis [AU]&151.51 & 1.42\\
Eccentricity&0.26& 0.3$\pm$0.04 \\
Inclination&114.1$^\circ$ &\\
Ascending node&84.13$^\circ$&\\
Argument of pericenter&247.74$^\circ$&41$\pm$8$^\circ$\\
Period&1420 years&535.7$\pm$3.1 days\\
\hline 
\end{tabular}
\tablecomments{For the distant star we used Keplerian elements published by \citet{1994AJ....107..306H}. For our calculations we used the orbital elements published by \citet{2004A&A...415..391M}. We used the planet's semi-major axis value 1.4273 AU for the calculations. The mass of the parent star HD19994 is $1.34 M_{Sun}$.}
\label{table:3} 
\end{table*}
%___________________________________________________
\begin{table*}[t]
\centering
\caption{Our proposal for possible orbital elements for planet HD19994 b.} 
\centering 
\begin{tabular}{l c c} % 3 columns 
\hline 
  &Prograde orbit&Retrograde orbit\\
\hline \hline 
Mass [$M_{Jup}$]&1.86$\pm$0.045 &1.86$\pm$0.045 \\
Semi-major axis [AU]&1.427&1.427\\
Eccentricity&0.300$\pm$0.04&0.300$\pm$0.04 \\
Inclination&65$^\circ \pm$3$^\circ$&115$^\circ \pm$3$^\circ$ \\
%Ascending node&264$^\circ \pm$4$^\circ$&84$^\circ \pm$4$^\circ$\\
Argument of pericenter&41$^\circ\pm$8&41$^\circ\pm$8\\
Period [day]&535.7$\pm$3.1&535.7$\pm$3.1\\
\hline 
\end{tabular}
\label{table:4} 
\end{table*}
%_________________________________________________________
The binary system HD 19994 (94 Ceti, ADS 2406 AB) contains an A component: a yellow-white dwarf with a mass of 1.34 $M_{Sun}$ and a B component: a red dwarf with a mass of 0.37 $M_{Sun}$. For the distant star we used Keplerian elements published by \citet{1994AJ....107..306H}. To define its semi-major axis $a_2$ we used two methods. Firstly, we derived this value using stellar parallax, where the distance was 22.38 pc \citep{2011A&A...532A..79S} and $6.77^{\prime\prime}$ for the angular separation \citep{1994AJ....107..306H} and resulted in a semi-major axis of 151.51 AU. Secondly, with the application of Keplerian law, using the values from Table \ref{table:3}, we calculated the value to be 151.37 AU. The difference between these two values is negligible we applied the value $a_2 = 151.51$ AU in the following calculations.

The planet HD 19994 b was discovered in 2000 \citep{2001Msngr.105....1Q} and is orbiting the A component. The taxonomy class is 2J0.2G3 \citep{2012AsBio..12..361P} and the minimum mass is 1.68 $M_{Jup}$. This planet orbiting with a semi-major axis of 1.42 AU with quite a high eccentricity $0.3 \pm 0.04$. For our calculations we used the orbital elements published by \citet{2004A&A...415..391M}. In Table \ref{table:3} the initial Kepler orbital elements for the planet and its distant star are shown. 

As with the first system, we revised the calculation for the semi-major axis of the planet using Keplerian law. With the values listed in Table \ref{table:3}, we valued the planet's semi-major axis at $a_1$=1.427 AU. If we varied the mass of the planet to 5 $M_{Jup}$ we would get a value of 1.428 AU. We decided to use the value 1.427 AU for the semi-major axis and 1.68 $M_{Jup}$ for the mass of the planet in our calculations.

As in the case of 16 Cyg b we varied the elements $I_1$ from $0^\circ$ to $180^\circ$, $1^\circ$ at a time  and $\Omega_1$ from $0^\circ$ to $360^\circ$, also for each degree. We recognized the value of the planet's inclination $I_1$ for which the maximum value of the planet's eccentricity $e_1$ is close to the initial value 0.300. 

According to our calculation we received the value \mbox{$a_1=1.427$ AU} 
 for the semi-major axis, $1.86 M_{Jup}$ for the mass, and $65^\circ$ or $115^\circ$ for the inclination of the planet. These values are presented in Table \ref{table:4} and were used for our next calculations. 

The results of our calculations are presented in Figure \ref{fig4}. and \ref{fig6}. If  the ascending node is $\Omega_1 \in \left( 146^\circ , 177^\circ \right) $ or $\Omega_1 \in \left( 350^\circ , 383^\circ \right) $ for prograde motion, $I_1=65^\circ$, and $\Omega_1 \in \left( 171^\circ , 202^\circ \right) $ or $\Omega_1 \in \left( 325^\circ , 358^\circ \right) $ for retrograde motion, $I_1=115^\circ$, the planet reaches the Roche limit in its pericentre ($d_R =0.052$ AU) with an eccentricity of $e_1 \ge 0.963$. In such a pericenter, large perturbations and tidal forces drastically effect the planet and lead to its destruction. 

For the values
$I_1=65^{\circ}$  and $\Omega_1=263^{\circ}$ the maximum and minimum values of the eccentricity $e_1$ are equal to the initial value $e_1 = 0.300$. In this case $q_0 =-0.999 < q_{01}=-0.739$. 
With these elements, the planet's orbit would be stable. 
%%%%%%%%%%%%%%%%%%%%%%%%%%%%%%%%
\section{Comparison of our theoretical results with numerical integration}
%_________________________________________________
\begin{figure}[t]
  \centering 
   \includegraphics[width=7.5cm]{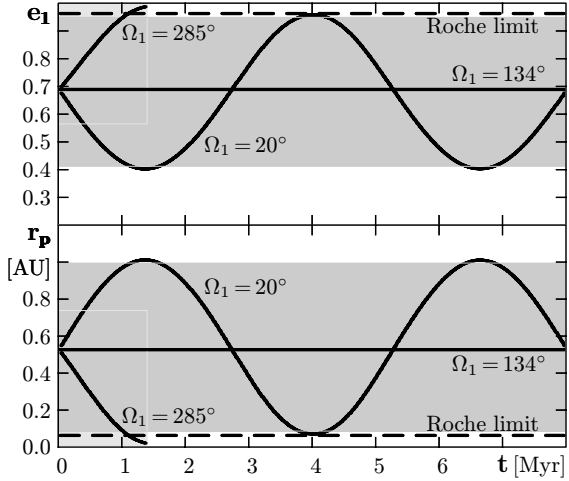}
     \caption{The planet 16 Cyg Bb. The evolution of the planet's eccentricity $e_1$ and the pericenter distance $r_p$  within the interval of $8\times10^6$ years. The curves are the result of the numerical integration. For the case if $I_1=45^{\circ}$  and $\Omega_1=20^{\circ}$ or $I_1=45^{\circ}$ and $\Omega_1=134^{\circ}$, the planet's orbit does not reach the Roche limit. For the case if $I_1=45^{\circ}$  and $\Omega_1=285^{\circ}$, the planet stays within the Roche limit. The boundaries of the gray zones for all three cases were computed from the our theory. For the case of $\Omega_1=134^{\circ}$, the gray zone is comparable with the line. The Roche limit is plotted by a dashed line.}
 \label{fig5}
 \end{figure}
%_______________________________________________
For confirmation of the obtained analytic results, we compared them with the results of numerical integration.  The equations of the motion of the systems were numerically integrated from the initial  date 2000 January 1, forward within the interval of $8 \times 10^6$ years, using the Everhart's integrator \citep{1985dcto.proc..185E}. We have  integrated the three variants of $\Omega_1$ for the prograde and retrograde orbits for both systems.

For 16 Cyg Bb we used  the initial value of the planet's inclination $i_1=45^{\circ}$ and the planet's ascending node $\Omega_1=20^{\circ}$, $\Omega_1=134^{\circ}$, and $\Omega_1=285^{\circ}$.
The results obtained by the theory and by the numerical integration agree quite well. For an illustration, the evolution of the eccentricity $e_1$ and the pericenter distance $r_p=a_1(1-e_1)$ of the planet 16 Cyg Bb are presented in Figure \ref{fig5}.

\begin{figure}[t]
   \centering 
   \includegraphics[width=7.5cm]{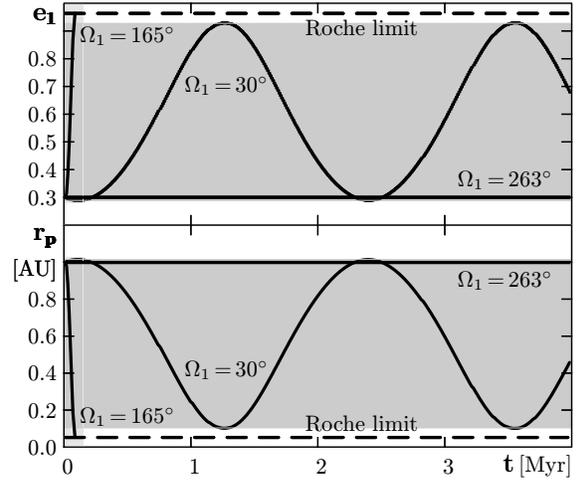}
     \caption{The planet HD19994 b. The evolution of the planet's eccentricity $e_1$ and the pericenter distance $r_p$  within the interval of $4\times10^6$ years. The curves are the result of the numerical integration. For the case if $I_1=65^{\circ}$  and $\Omega_1=30^{\circ}$  or $I_1=65^{\circ}$  and $\Omega_1=263^{\circ}$, the planet's orbit does not reach the Roche limit. For the case $I_1=65^{\circ}$  and $\Omega_1=165^{\circ}$, the planet stays within the Roche limit. The boundaries of the gray zones for all  three cases were computed from the our theory. For the case of $\Omega_1=263^{\circ}$, the gray zone is comparable with the  line. The Roche limit is plotted by a dashed line.} 
  \label{fig6}
\end{figure}

For HD 19994 b (see Figure \ref{fig6}) we used as the initial values $i_1=65^{\circ}$ and $\Omega_1=30^{\circ}$, $\Omega_1=165^{\circ}$, and $\Omega_1=263^{\circ}$. As in the case of 16 Cyg Bb, the results of the numerical integration are the same as the results we obtained by the analytical theory.

We obtained identical results from the numerical integration and analytical theory for the retrograde orbits of the both systems, too.
%_______________________________________________
%%%%%%%%%%%%%%%%%%%%%%%%%%%%%%%%%%%%%%%%%%%%%%%%%%%%%%%%%%
\section{Comparison with previous works}
In our investigation we have used the methods of the classical celestial mechanics, developed by Hamilton. We come from the Principle of Determinacy (PD)\citep{Lidov} , that if the initial conditions of each object in a mechanical system are defined at any instant of time, then their further behavior is defined expressly. 

Note, that Hamilton's equations are first order in the time derivative, which makes them more convenient for computation. The Hamiltonian (\ref{Hamiltonian}) of the system (\ref{canonical}) and (\ref{equationmotion}), without short-periodic terms, permits the solution in which the secular and the long-periodic terms are taken into account in the intermediate orbit and allows the close approach of the EP to the star to be established. 

The Long-periodic stability of planets in a binary system was investigated by \citet{1999AJ....117..621H}. The planets are taken to be test particles moving in the field of an eccentric binary system. This study investigated the orbital stability numerically, with the elliptic restricted three body problem. We made a comparison with their results concerning planet \mbox{16 Cyg Bb}. We used the elements of this planet for the application of our theory. \citet{1999AJ....117..621H} integrated differential equations of the motion using $\Omega_1=0^\circ$ and obtained instability between $10^7$ and $10^9$ years. We varied this element from $0^\circ$ till $360^\circ$ and obtained instability around $\Omega_1=0 ^\circ$, too. 

The secular dynamics of massless particles orbiting a central star and perturbed by a secondary star component with high eccentricity have been investigated \citet{2011A&A...530A.103G}. They used the Lie series perturbation scheme restricted to second order in the small parameter. Their results have shown that the second-order secular dynamic reproduces the behaviour of a planet with good precision.

We eliminated the short-periodic terms by von Zeipel's method in the general three-body problem. This method allows us to estimate values for the short periodic terms; these values are less than $\pm 10^{-3}$. The short-periodic terms are small and do not influence the evolution and stability. Our condition of stability is valid at any time interval. Our theory may also be used for EPs with a large mass.

In the Sun--asteroid--Jupiter problem, the Hamiltonian describing the motion of the massless asteroid in the heliocentric coordinates was used by \citet{1997AJ....113.1915I}. They used the so-called Kozai mechanism which shows the behaviour of the eccentricity as a function of the initial inclination. In the restricted three-body problem, with the averaging over the mean anomalies, the perturbation function there are the integrals:
\begin{eqnarray}
\sqrt{1-e^2} \sin i=const.
\nonumber
\end{eqnarray}
and
\begin{eqnarray}
e^2 \left(\frac{2}{5}-\sin^2i \sin^2\omega \right)=const.
\end{eqnarray}
These were obtained at nearly the same time by \citet{1962P&SS....9..719L} and \citet{1962AJ.....67..591K}. The integrals permit us to execute qualitative analysis of the family of the phase trajectories.

In our case, the connection between the eccentricity, the inclination, and the argument of pericenter is more complex, depends on all Keplerian elements, and defined by the equation (\ref{hh}). It allows us to arrive at the some conclusion, similar to the results of \citet{1997AJ....113.1915I}. The increase of the eccentricity occurs even if the third body is very distant and the perturbation is small. 

The time scale needed to observe this phenomenon is quite lengthy, for example, the sudden increase in the eccentricity of Neptune is observed at 100 Myr. \citep{1997AJ....113.1915I}. But what is clear is that instability only occurs when the mutual inclination of a planet's orbit and the distant star's orbit is high. So we suppose, for the stability of a planet's orbit, the conditions are demanded: the initial value of the cosine of the mutual inclination $q_0 < q_{01}$ or $q_0 >q_{02}$, where $q_{01}$ and $q_{02}$ are roots of expression (\ref{q}) and in this case $ \overline{c}-\overline{G}_2>0$.  
%%%%%%%%%%%%%%%%%%%%%%%%%%%%%%%%%%%%%%%%%%%%%%%%%%%%%%%=
\section{Conclusion}
We have shown that an EP revolving in a binary system around one of the components, may move on a stable or unstable orbits. The conditions for a stable motion depend on the orbital parameters, which can be calculated from the formulas of the previous sections. There are the angle of the mutual inclination between orbits, the angular momentum of the distant star and the maximum value of the eccentricity of the planet's orbit. When the value of the planet's eccentricity is close to one, in the pericenter the planet reaches the Roche limit. The tidal forces of the star destroy the planet. We have suggested the possible values for the unknown elements with which the orbit of the planet would remain stable. 

The results of our calculations are presented in Figures 1--6. The results of the numerical integration support the results obtained by the analytical theory.

For 16 Cyg Bb, there are three possible regions of stability. Firstly, for the prograde orbit, $I_1=45^\circ$, and the ascending node $\Omega_1 \in \left[ 14^\circ , 249^\circ \right]$. Secondly, for the retrograde orbit, $I_1 =135^\circ$, and the ascending node $\Omega_1 \in \left[ 0^\circ , 74^\circ \right]$ or  $\Omega_1 \in \left[ 198^\circ , 360^\circ \right]$.  We proposed the planet's mass to be 2.38$\pm$0.04 $M_{Jup}$ and the value of its semi-major axis equal to 1.693 AU, for which the third Keplerian law is valid.

For the second system HD19994, the values of inclination and the ascending node of the planet for which the motion is stable are: for the prograde motion, $I_1=65^\circ$, and the ascending node 
$\Omega_1 \in \left[ 23^\circ , 146^\circ \right]$ or $\Omega_1 \in \left[ 177^\circ , 350^\circ \right]$; for the retrograde motion, $I_1=115^\circ$, and the ascending node 
$\Omega_1 \in \left[ 202^\circ , 325^\circ \right]$ or $\Omega_1 \in \left[ 358^\circ , 0^\circ \right]$ or $\Omega_1 \in \left[ 0^\circ , 171^\circ \right]$.
For the planet's mass, we proposed the value of 1.86$\pm$0.045 $M_{Jup}$ and for the planet's semi-major axis, a value of 1.427 AU. These values are in accordance with the third Keplerian law.
%%%%%%%%%%%%%%%%%%%%%%%%%%%%%%%%%%%%%%%%%%%%%%%%%%%%%%%%%%%%%%%%%%
\acknowledgments
This work was supported by the Slovak Research and Development Agency under the contract No. APVV-0158-11.
%%%%%%%%%%%%%%%%%%%%%%%%%%%%%%%%%%%%%%%%%%%%%%%%%%%%%%%%%%%%%%%%%%

%%%%%%%%%%%%%%%%%%%%%%%%%%%%%%%%%%%%%%%%%%%%%%%%%%%%%%%%

\end{document}